\documentclass[10pt,twocolumn,letterpaper]{article}

\usepackage[pagenumbers]{cvpr} 
\definecolor{cvprblue}{rgb}{0.21,0.49,0.74}
\usepackage[pagebackref,breaklinks,colorlinks,allcolors=cvprblue]{hyperref}

\title{VerbDiff: Text-Only Diffusion Models with Enhanced Interaction Awareness}

\author{SeungJu Cha \quad Kwanyoung Lee \quad Ye-Chan Kim \quad Hyunwoo Oh \quad Dong-Jin Kim\\
  Hanyang University, South Korea \\ 
    \tt\small{\{sju9020, mobled37, dpcksdl78, komji, djdkim\}}@hanyang.ac.kr 
    }

\begin{document}
% \maketitle
\twocolumn[{
\renewcommand\twocolumn[1][]{#1}
\maketitle
\begin{center}
    % \vspace{-15pt}
    \captionsetup{type=figure}
    \includegraphics[width=0.98\linewidth]{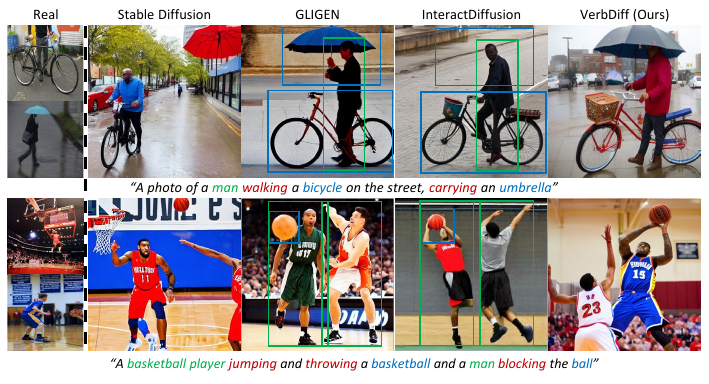}
    \captionof{figure}{\textbf{Generated samples illustrating multiple human-object interactions.} Each color represents distinct humans, objects, and interaction words. GLIGEN~\cite{li2023gligen} and InteractDiffusion~\cite{hoe2024interactdiffusion} use grounding boxes as additional conditions, whereas Stable Diffusion~\cite{rombach2022high} and VerbDiff rely solely on text.}
    \label{fig:teaser}
\end{center}
}]
\begin{abstract}
\label{sec:abstract}
Recent large-scale text-to-image diffusion models generate photorealistic images but often struggle to accurately depict interactions between humans and objects due to their limited ability to differentiate various interaction words.
In this work, we propose VerbDiff to address the challenge of capturing nuanced interactions within text-to-image diffusion models. VerbDiff is a novel text-to-image generation model that weakens the bias between interaction words and objects, enhancing the understanding of interactions. 
Specifically, we disentangle various interaction words from frequency-based anchor words and leverage localized interaction regions from generated images to help the model better capture semantics in distinctive words without extra conditions. 
Our approach enables the model to accurately understand the intended interaction between humans and objects, producing high-quality images with accurate interactions aligned with specified verbs. Extensive experiments on the HICO-DET dataset demonstrate the effectiveness of our method compared to previous approaches.
\end{abstract}    
\section{Introduction}
\label{sec:intro}
Recently, text-to-image diffusion models have demonstrated the ability to produce photorealistic images from natural texts~\cite{rombach2022high}.
However, these models sometimes struggle to accurately depict interactions between humans and objects~\cite{hoe2024interactdiffusion}.
For example, they often fail to differentiate between semantically distinct prompts, such as ``A person \textit{walking} a bicycle'' and ``A person \textit{riding} a bicycle,'' particularly in capturing the semantics in verbs that are crucial for accurately depicting the intended interaction.
 
We hypothesize that this problem arises from the strong object bias of CLIP~\cite{radford2021learning}, which fails to adequately capture semantic differences between verbs~\cite{momeni2023verbs}.
Specifically, CLIP~\cite{radford2021learning} often focuses on objects in prompts, overlooking semantic distinctions in interaction verbs.
To enhance prompt interpretation ability of models, previous methods have leveraged additional modules such as large language models (LLMs)~\cite{gani2023llm, lian2023llm, feng2024layoutgpt} or extra bounding boxes~\cite{li2023gligen, xie2023boxdiff,zheng2023layoutdiffusion, hoe2024interactdiffusion} to provide explicit relational visual information.

\begin{figure}
    \centering
    \includegraphics[width=1.0\linewidth]{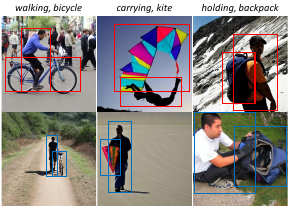}
    \caption{\textbf{Examples of interactions from InteractDiffusion~\cite{hoe2024interactdiffusion}, showing a lack of understanding of interaction words.} The model relies on precise bounding boxes rather than understanding interaction words to exhibit accurate interactions, as shown in the results for the “red” and “blue” boxes.}
    \label{fig:introfig1}
\end{figure}

For example, InteractDiffusion~\cite{hoe2024interactdiffusion} introduces controllable image generation using Human-Object Interaction (HOI) information with bounding boxes for human, relation, and object $<$h, r, o$>$ triplets.
Nevertheless, as shown in Fig.~\ref{fig:introfig1}, the model still struggles with accurate interactions even with these additional conditions. 
For instance, given the prompt “walking, bicycle,” the model fails to depict the intended interaction when semantically distinctive interaction words share a similar bounding box (\eg \textit{walking} and \textit{riding}). 
This suggests that the model lacks an understanding of interaction semantics and relies heavily on precise bounding boxes, which are labor-intensive to provide.

To improve the model's interaction understandability without additional conditions, we propose VerbDiff, a 
novel text-to-image generation method that better distinguishes interaction words via several objectives.
First, we apply Relation Disentanglement Guidance (RDG) using frequency-based anchor texts from the dataset for each human-object pair to reduce interaction bias in generated images.
We observe that the exhibited interactions in generated images have bias, which follows frequent verbs corresponding to each human-object pair in the data distribution.
To address this, we align interaction features between real and generated images while separating those of generated images from anchor text features, enabling a more effective distinction of interaction words.
Additionally, we introduce a balancing factor to address the extreme long-tail distribution of interaction words across human-object pairs, adjusting interaction modification during training.
This allows the model to learn diverse interactions in a balanced way.

Secondly, to 
enforce the model to better focus on
the interaction regions between humans and objects, we introduce Interaction Direction Guidance (IDG).
This approach emphasizes fine-grained semantic distinctions in localized interactions.
We design an Interaction Region (IR) module to capture specific interaction areas in generated images, leveraging cross-attention maps in text-only diffusion models.
The IR module extracts interaction regions using the centroids of cross-attention maps associated with $<$h, r, o$>$ tokens.
We obtain biased interaction features from these regions and apply interaction direction guidance.
This guidance effectively modifies interaction regions within images, guiding Stable Diffusion (SD) to concentrate more accurately on these regions while capturing nuanced interaction details. 
We validate our method's effectiveness on the HICO-DET dataset~\cite{chao2018learning}.
Our approach accurately captures semantic differences between interaction words and generates high-quality images with accurate interactions compared to SD~\cite{rombach2022high}, InteractDiffusion~\cite{hoe2024interactdiffusion}, and other prior methods.
Additionally, we introduce a new metric using Sentence BERT~\cite{reimers2019sentence} for image generation tasks.
This metric offers improved capability in discerning subtle and nuanced meanings of interaction words compared to CLIP.
By focusing on semantic distinctions in interactions, this metric enables more precise evaluation, complementing CLIP similarity. 
Our contributions are as follows:

(1) We introduce VerbDiff, a novel text-to-image model that enhances the semantic interaction understanding by distinguishing interaction verbs and generating images with accurate human-object interactions without any additional conditions. (2) To reduce generating images with biased interactions, we apply relation disentanglement guidance using frequency-based anchor text extracted from real data distributions. (3) Our proposed IR module captures detailed interaction regions using cross-attention map centroids, providing specified-interaction region guidance without explicit bounding boxes. (4) We implement a Sentence-BERT metric to effectively capture semantic differences between interaction sentences, addressing CLIP's limitations. Extensive experiments demonstrate that our approach outperforms previous methods across various settings.
\section{Related Work}
\label{sec:related}

\subsection{Diffusion Models}
Recently, diffusion models have achieved state-of-the-art performance in text-to-image generation~\cite{rombach2022high} and controllable image generation~\cite{hoe2024interactdiffusion, li2023gligen, zhang2023adding}.
While Stable Diffusion (SD)~\cite{rombach2022high} integrates conditioning embeddings into cross-attention layers for text-to-image generation, it often struggles to fully align generated images with text instructions.
This arises from inherent limitations of texts to convey exact visual information and strong object bias of CLIP~\cite{momeni2023verbs}.

To address these, recent research has focused on controllable image generation using bounding boxes~\cite{xie2023boxdiff,li2023gligen,zheng2023layoutdiffusion}.
For example, GLIGEN~\cite{li2023gligen} extends SD to incorporate grounding inputs such as bounding boxes, image prompts, key points, and spatially aligned conditions.
LayoutDiffusion~\cite{zheng2023layoutdiffusion} enables multiple bounding box inputs for layout-specific generation.
Although these methods better generate images, they require additional conditions beyond text input to achieve the intended image outputs.
In contrast, our proposed method achieves comparable results without additional grounding information, providing a more efficient alternative for precise and flexible generation.

\subsection{Relation Generation}
Relation generation between objects has been explored leveraging LLMs~\cite{gani2023llm, lian2023llm, feng2024layoutgpt} and scene graph~\cite{wu2024relation, xu2024joint, yang2022diffusion, wu2024imagine, liu2024r3cd, shen2024sg}.
While LLM-based methods can generate accurate object layouts, they primarily focus on compositional relations rather than interactions~\cite{gani2023llm, lian2023llm, feng2024layoutgpt}.
Scene graph-based approaches have addressed these limitations by combining scene graph encoders to enhance prompt understandability~\cite{wu2024relation, xu2024joint, yang2022diffusion, wu2024imagine, liu2024r3cd, shen2024sg}. 
However, they require additional training to extract scene graphs from prompts and struggle to capture subtle semantic differences in interactions.

To depict accurate interactions during generation, researchers have focused on human-object interactions (HOI)~\cite{jia2024customizing, huang2023reversion, jiang2024record, hoe2024interactdiffusion}.
Some methods use inversion-based frameworks to capture interaction semantics but require optimization for each interaction word~\cite{huang2023reversion, jia2024customizing}.
More generalized approaches leverage additional conditions such as human poses~\cite{jiang2024record} or bounding boxes~\cite{hoe2024interactdiffusion}.
Specifically, InteractDiffusion~\cite{hoe2024interactdiffusion} leverages the HOI information with bounding boxes to generate interactions. 
However, this method still relies on additional information rather than semantic differences between prompts.
Our work is similar to InteractDiffusion~\cite{hoe2024interactdiffusion} by leveraging HOI information but differs from focusing on semantic differences between interaction words to enhance interaction understanding in SD.
This approach aims to improve the generation of nuanced interactions without relying on additional conditions. It could potentially lead to more accurate and diverse depictions of human-object interactions in synthesized images.
\section{Method}
\label{sec:method}
\begin{figure*}[t]
    \centering
    \includegraphics[width=1.0\linewidth]{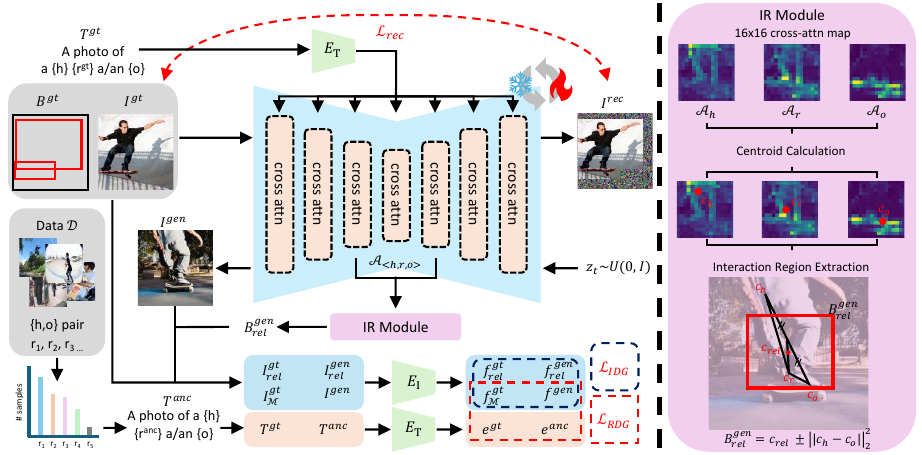}
    \caption{\textbf{Pipeline of VerbDiff.} VerbDiff uses Relation Disentanglment Guidance (Sec.~\ref{subsec:RDG}) that separates the interaction features from the anchor text for each human-object pair. Additionally, it contains the IR module (Sec.~\ref{subsec:IDG}) (right) which extracts localized interaction regions from generated images without explicit bounding boxes and Interaction Direction Guidance  (Sec.~\ref{subsec:IDG}) that guides the model to focus more on the fine-grained interaction regions.}
    \label{fig:Verbdiff}
\end{figure*}

Our goal is to generate images with accurate interactions by enhancing the model understandability of the interaction word without additional conditions. 
Our proposed method, VerbDiff as illustrated in Fig.~\ref{fig:Verbdiff} contains two main parts:
(1) Relation disentanglement guidance leverages frequency-based anchor texts from human and object pairs along with input texts and real and generated image features.
This guidance helps visually align the generated interaction features with real ones and distinguish the semantic differences between interaction words. 
(2) Interaction direction guidance that helps the model concentrate more on the interaction region while modifying the image features.
To capture the interaction region in arbitrarily generated images, we further design an interaction region module (IR module) to extract a centroid-based interaction region from the cross-attention map corresponding to $<$h,r,o$>$ tokens.

\subsection{Preliminary}
\noindent\textbf{SD Training Objective}. 
Text-to-image generation models, such as Stable Diffusion (SD)~\cite{rombach2022high}, show photo-realistic images that align with given text prompts through an efficient two-stage training scheme. 
The first stage learns to map the input images $I$ into latent space $z$. 
The second stage learns the denoising UNet to gradually predict the real noise in the diffusion forward process.
Starting from the original latent image $z_0$, the diffusion model gradually adds noise $\epsilon$ to produce noisy sample $z_t$ through uniformly sampled time steps $t$. 
Then, with text condition $T$, the UNet predict the noise $\epsilon_\theta$ for each time step to reconstruct $z_t$ to $z_0$  as follows:
\begin{align}
\underset{\theta}{\min}~\mathcal{L}_{\text{LDM}} = \mathbb{E}_{z, \epsilon \sim \mathcal{N}(0, 1), t} \left[ \left\lVert \left( \epsilon - \epsilon_{\theta}(z_t, t, T) \right) \right\rVert_2^2 \right].
\label{eq:recon}
\end{align}

\noindent\textbf{Cross-Attention Layer}.
When given text, the CLIP~\cite{radford2021learning} text encoder produces text embedding, and the UNet in SD takes embedding and computes cross-attention between encoded text $e$ and intermediate image features $\phi(z_t)$:
\begin{align}
    % Q_{\text{cross}} &= W_Q\phi(z_t),\quad K_{\text{cross}} = W_Kc_{emb},\\
    \mathcal{A} &= \text{Softmax} \left( \frac{Q K^{\top}}{\sqrt{d} }\right),
\end{align}
where query $Q=W_Q \cdot \phi(z_t)$ and key $K=W_K \cdot e$ are computed with projection matrix $W_Q$, $W_K$ and $d$ is the dimension of $Q$. 
These cross-attention maps $\mathcal{A}$ 
enable the text-to-image diffusion models to depict texts into image features.
In addition, each $\mathcal{A}_j$, a cross-attention map corresponding to $j^{th}$ token in prompts, represents the attention weight on intermediate image features in various resolutions.

\subsection{Relation Disentanglement Guidance}
\label{subsec:RDG}
To depict accurate interaction in generated images without additional conditions, we first focus on the semantic difference between interaction words, especially the verbs. 
As the SD is known to have limitations in capturing the semantic nuances in interaction words, we apply relation disentanglement guidances.
The goal of this guidance is to alleviate
the relation bias within human-object pairs and improve interaction comprehension through anchor text, input text, and real and generated images.

First, given the ground-truth text $T^{gt} =$ ``\textit{A photo of a} $\{h\}$ $\{r^{gt}\}$ \textit{a/an} $\{o\}$'' (\eg ``A photo of a person \textit{holding} a backpack''), SD generates images $I^{gen}$ during training.
Although $T^{gt}$ specifies the target interaction, $I^{gen}$ often contains inaccurate and biased interactions that fail to meet verbs between each human and object pair.
We define a sentence with accurate interaction verbs corresponding to inaccurate interaction image $I^{gen}$ as $T^{anc}$.

In particular, we observe biased interaction images tend to depict verbs that appear more frequently in the data along human-object pair.
For instance, when generating images with the object ``backpack,'' the images commonly show a person \textit{wearing} a backpack, which is the most frequent verb in the dataset.
Based on this observation, we hypothesize that the biased interaction images typically align with frequent verbs in the real data distribution. 
Therefore, we define $r^{anc}$ as the verb with the highest sample count for each human-object pair as follows:
\begin{align}
r^{anc} = \underset{r \in R_o }{\arg\max} \, ~\mathcal{C}(r|o),
\end{align}
where $R_o$ denotes the set of verbs related to each human-object pair in dataset $\mathcal{D}$ and $\mathcal{C}(\cdot)$ represents the number of samples corresponding to each verb $r$ when given $o$.

Given the anchor verb $r^{anc}$, to reduce the generation of biased interaction images, we encourage SD to create images that with distant semantics from $T^{anc}=$ ``\textit{A photo of a} $\{h\}$ $\{r^{anc}\}$ \textit{a/an} $\{o\}$'' (\eg ``\textit{wearing} a backpack'').
Specifically, we guide the $I^{gen}$ closer to the accurate interaction sentence $T^{gt}$ while pushing away from the sentence with anchor interaction $T^{anc}$ by leveraging triplet loss~\cite{hoffer2015deep}. 

We extract normalized feature $f^{gen}$, $e^{gt}$, and $e^{anc}$ through CLIP~\cite{radford2021learning} encoder $E_T$ and $E_I$ from 
$I^{gen}$, $T^{gt}$, and $T^{anc}$, respectively.
Then, we set $e^{gt}$ as positive and $e^{anc}$ as negative when given anchor $f^{gen}$
and compute the cosine similarity with the generated image features, $f^{gen}$, and each text feature, $e^{gt}$ and $e^{anc}$.
Finally, we introduce a triplet loss to maximize the cosine similarity between $f^{gen}$ and $e^{gt}$ larger than the similarity between $f^{gen}$ and $e^{anc}$ with a margin of $m$ as follows:
\begin{align}
\mathcal{L}_{\text{triple}} = \max\left(0, m + \text{sim}\left(f^{gen}, e^{gt}\right) - \text{sim}\left(f^{gen}, e^{anc}\right)\right).
\end{align}

In addition to distinguishing the semantic difference between verbs, to generate accurate interaction at the image level as well, we add image aligning loss to make the model follow the interaction in $I^{gt}$.
However, $I^{gt}$ often contains multiple humans, objects, and interactions within a single image 
while we only use a single human, object, and interaction triplet as input during image generation.
This input discrepancy makes it difficult for the model to focus solely on the relevant interaction specified by the input prompts, causing confusion when depicting the specified interaction words.
To focus on the part in the image corresponding to the interaction label, we extract mask region $\mathcal{M}$ combining the human and object bounding boxes in a dataset and obtain $I^{gt}_\mathcal{M} = I^{gt} \odot \mathcal{M}$.
Then, we extract $f^{gt}_\mathcal{M}$ through image encoder and minimize the cosine similarity between $f^{gt}_\mathcal{M}$ and $f^{gen}$ as bellow:
\begin{align}
% \mathcal{L}_{\text{align}} = 1 - \frac{}{}
\mathcal{L}_{\text{align}} = 1 - \frac{f^{gt}_\mathcal{M} \cdot f^{gen}}{|f^{gt}_\mathcal{M}||f^{gen}|}. 
\end{align}

\noindent\textbf{Adaptive Interaction Modification.} 
When applying $\mathcal{L_{\text{triple}}}$ and $\mathcal{L_{\text{align}}}$ to capture the semantic differences between interaction words,
we find that the modification extent applied to the generated images differs across interaction words.
As each interaction text $T^{gt}$ has widely varying sample counts, we also focus on balancing the modification extent across interaction words.
Specifically, we adaptively scale the loss considering the number of samples following~\cite{cui2019class}.
We apply an effective number, which is defined as:
\begin{align}
\alpha(k)= \frac{1-\beta^{n_k}}{1-\beta}, 
\end{align}
where $\beta=(N-1)/N$, $n_k$ is the number of samples in class $k$, and N is the total number of samples in the dataset.

\noindent\textbf{Relation Disentanglement Guidance.} Finally, we multiply $\alpha$ on $\mathcal{L_{\text{triple}}}$ and $\mathcal{L_{\text{align}}}$, defining relation disentanglement guidance as follows:
\begin{align}
\mathcal{L}_{\text{RDG}} = \alpha\cdot(\mathcal{L_{\text{triple}}} + \mathcal{L_{\text{align}}}).
\end{align}
This guidance induces SD to generate images with accurate interactions for each verb and to distinguish the semantic difference between interactions.

\subsection{IR Module $\And$ Interaction Direction Guidance}
\label{subsec:IDG}

Relation Disentanglement Guidance (RDG) effectively enables the model to better distinguish and represent each interaction word, helping to exhibit the intended interactions at the image level. 
However, the model sometimes shows images that misalign with human expectations, especially in the local regions where human-object interactions occur. We hypothesize that focusing on finer, localized interaction regions helps the model improve the detailed interaction expressions, thereby allowing the model to better capture more region-specific interaction information.

To focus more on specific interaction regions, we apply interaction direction guidance (IDG), leveraging interaction regions extracted from the interaction region module with cross-attention maps corresponding to the $<$h,r,o$>$ tokens in generated images. 
IDG enhances the model's ability to depict region-level details, guiding the depicted interactions to align more perceptually close to human intentions.

\noindent\textbf{Interaction Region Module.} 
To achieve a more detailed interaction expression, we first extract interaction regions from both real and generated images.
Although real images contain explicit human and object bounding boxes in the dataset, it is challenging to extract specific interaction regions without additional conditions from generated images.
Recently,~\cite{hertz2022prompt} demonstrated that the cross-attention maps capture the human or object's existence corresponding to each token in prompts.
We leverage the cross-attention maps $\mathcal{A}_h$, $\mathcal{A}_r$ and $\mathcal{A}_o$ corresponding to $<$h,r,o$>$ token to extract interaction region between human and objects.

However, it is nontrivial to directly extract the specific interaction region from a continuous attention map. 
We apply the centroid extraction mechanism proposed in~\cite{epstein2023diffusion} to compute specific points within attention maps. 
Specifically, we calculate $c_h$, $c_r$ and $c_o$ from each cross-attention maps as follows:
\begin{align}
c = \frac{1}{\sum_{h,w} \mathcal{A}} 
\begin{bmatrix}
\sum_{h,w} w \cdot \mathcal{A} \\
\sum_{h,w} h \cdot \mathcal{A}
\end{bmatrix},
\end{align}
where $h$ and $w$ denote the height and width in the attention map $\mathcal{A}$.
Then, we define the interaction center $c_{rel}$ as the centroid of a triangle defined by $c_h$, $c_r$, and $c_o$.
Finally, we extract interaction region $B_{rel}^{gen}$ by leveraging the distance between human and object center with $B_{rel}^{gen} = c_{rel}  \pm  \|c_h - c_o\|_2^2$. 
Additionally, we extract $B_{rel}^{gt}$ in the same manner, leveraging the human and object bounding boxes provided in the dataset. We treat $c_{rel}$ as the midpoint between the human and object centers from the given boxes $B^{gt}$.

\noindent\textbf{Interaction Direction Guidance.} 
To generate images with more realistic interactions, we design guidance that aligns feature differences at the image level with those at the interaction region.
With interaction region $B_{rel}^{gt}$ and $B_{rel}^{gen}$, we obtain interaction region image by masking the real and generated images: $I_{rel}^g = B_{rel}^g \odot I^g,\quad \text{where } g \in \{gt, gen\}$.
We then encode both interaction region images through a CLIP image encoder and extract interaction region features $f_{rel}^{gt}$ and $f_{rel}^{gen}$.
Leveraging interaction region features, we design the interaction direction guidance to focus the model on modifying the image feature, especially within the interaction region.
Specifically, we calculate the direction between $f_{rel}^{gt}$ and $f_{rel}^{gen}$, referred to as biased relation feature $f_{rel}^{bias} = f_{rel}^{gt} - f_{rel}^{gen}$, to align the direction of biased interaction features with that of the real images.
The interaction direction guidance is as follows:
\begin{align}
\mathcal{L}_{\text{IDG}} = 1 - \frac{(f^{gt}_\mathcal{M} - f^{gen}) \cdot (f^{bias}_{rel})}{|f^{gt}_\mathcal{{M}} - f^{gen}| |f^{bias}_{rel}|}.
\end{align}

\subsection{Training Phase}
We train only the cross-attention layer in SD to capture the semantically distinct relation words, as the cross-attention reflects the existence of each word token~\cite{liu2024towards}.
We adopt the reconstruction loss (Eq.~\ref{eq:recon})
to train T2I models in addition to relation disentanglement guidance and interaction direction guidance.
However, when an image contains multiple humans and objects, it can lead to multiple interactions that do not match a single target interaction verb.
To accurately separate the human and object corresponding to interaction words, we apply mask $\mathcal{M}$ when calculating the reconstruction loss as follows:
\begin{align}
\mathcal{L}_{\text{rec}} = \mathbb{E}_{z, \epsilon \sim \mathcal{N}(0, 1), t} \left[ \left\lVert \left( \epsilon \odot \mathcal{M} - \epsilon_{\theta}(z_t, t, T) \odot \mathcal{M} \right) \right\rVert_2^2 \right].
\end{align}

Finally, we combine reconstruction loss with disentanglement guidance and interaction direction guidance leveraging  $\lambda_1$, $\lambda_2$, and $\lambda_3$ as below:
\begin{align}
\mathcal{L}_{\text{total}} = \lambda_1\cdot\mathcal{L}_{\text{rec}} + \lambda_2\cdot\mathcal{L}_{\text{RDG}} + \lambda_3\cdot\mathcal{L}_{\text{IDG}}.
\end{align}

\section{Experiments}
\label{sec:exp}

\begin{figure*}[t]
    \centering
    \includegraphics[width=0.9\textwidth, height=0.8\textwidth]{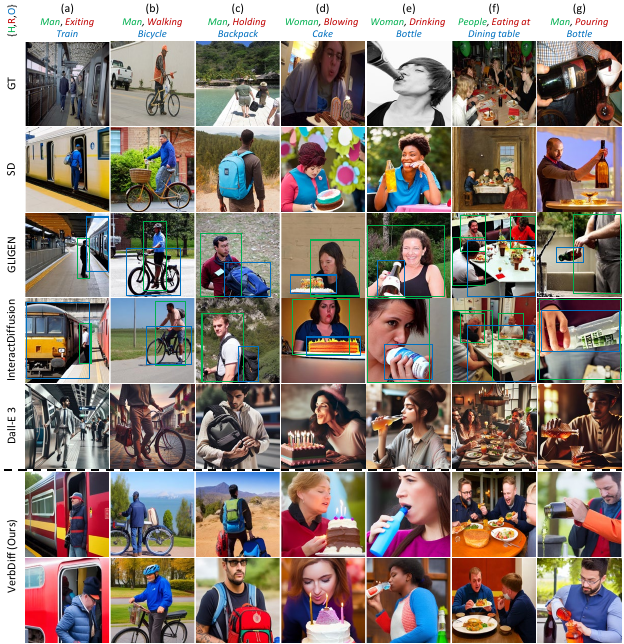}
    \caption{\textbf{Interaction comparison in generated images with other models.}  We generate images using a fixed template: ``A \{H\} \{R\} a/an \{O\}''. The top row displays the input human, interaction word, and object. Green and blue boxes represent additional grounding boxes used during image generation for each human and object, respectively. 
    Our results produce images with more accurate interactions than other models, closely resembling ground-truth interactions while maintaining high image quality comparable to large-scale generative models like DALL·E 3~\cite{betker2023improving}.}
    % Compared to other models, our results generate images with more accurate interactions, closely resembling ground-truth interactions while achieving high-quality images comparable to large-scale image generation models like Dall-E 3~\cite{betker2023improving}.}
    \label{fig:qual1}
\end{figure*}

We train our model on SD v1.4 at a resolution of $512 \times 512$. Using the Adam optimizer with a learning rate of $4 \times 10^{-6}$, we trained for a single epoch over 17 hours with a batch size of 12 on 3 Nvidia A6000 GPUs. For image generation during training, we utilize the DDIM scheduler~\cite{song2020denoising} with 30 sampling steps, and during inference, we increase sampling steps to 50. Finally, $\lambda_1$, $\lambda_2$, and $\lambda_3$ are set to 1.0, 10, and 0.8, respectively, and the margin $m$ is set to 0.2.
% More details are in supplementary materials.
% Appendix Sec.~\ref{sec:implement}.

\subsection{Dataset}
We conduct our experiments on the HICO-DET dataset~\cite{chao2018learning}, which consists of 47,776 images (38,118 for training and 9,658 for testing). The dataset provides annotations for 600 HOI categories using 80 object categories and 117 verbs.

To train our model, we realign the training set according to the 600 HOI text descriptions and modify annotations to match each image with its corresponding text descriptions. This realignment yields 69,404 images across 600 prompts. We exclude text classes containing ``and'' (\eg ``A photo of a person and an airplane'') to avoid ambiguity from various interactions within the same class, resulting in a final training set of 61,114 images across 501 prompts. 
More details about the dataset are in the supplementary materials.
% Appendix Sec.~\ref{sec:data}.

\subsection{Metrics}
\noindent{\textbf{CLIP$\And$Sentence-BERT Similarity.}}
We assess the model understandability of interaction words by calculating text-to-text similarity scores using both CLIP~\cite{radford2021learning} and Sentence-BERT (S-BERT)~\cite{reimers2019sentence} that we newly adopt to image generation task. After generating captions with InstructBLIP~\cite{liu2024visual}, we compute cosine similarity against ground-truth texts. 
We also measure the similarity between the input prompts and generated images through CLIP.
Details about the S-BERT metric are in the supplementary materials.
% Appendix Sec.~\ref{sec:sbert}.

\noindent{\textbf{HOI Classification Accuracy.}} 
To evaluate the robustness of the model understanding interaction words, we measure interaction accuracy with generated images using the SOV-STG~\cite{chen2023focusing} HOI detection model pre-trained on HICO-DET~\cite{chao2018learning}. 
We report scores on two different weights, SOV-STG-S and SOV-STG-Swin-L, with different backbones (\eg ResNet~\cite{he2016deep} and Swin-L~\cite{liu2021swin}).
Accuracy is reported in two main settings: Known Object (KO), which measures verb accuracy assuming the objects are correct, and Default (Def.), which assesses both verb and object correctness, with each setting further divided into \textit{Full} (all classes) and \textit{Rare} (classes with fewer than 10 samples).

\noindent{\textbf{VQA-Score.}}We additionally use VQAScore~\cite{lin2025evaluating}, a large image-text alignment benchmark for generative models, to evaluate comprehensive interaction alignment scores.
This score measures the model’s ability to understand more complex compositional prompts, such as spatial and action relationships. For assessing the accuracy of depicting interaction between humans and objects in images, we use two main metrics: the Image-to-Text (I2T) alignment score and the VQA score. The I2T alignment score improves similarity measurement between complex prompts containing interaction words and generated images by incorporating a large language model. The VQA score measures the probability that a VQA model will answer ``yes'' to questions containing interaction words paired with generated images. 
% More details are in the supplementary materials.
% Appendix Sec.~\ref{sec:VQA}.

\subsection{Qualitative Results}

Fig.~\ref{fig:qual1} shows a comparison of generated images focusing on the interactions between humans and objects across our method and other models. Compared to previous approaches, our results exhibit more detailed interactions, closely resembling ground-truth images and producing high-quality outputs comparable to images from DALL-E 3~\cite{betker2023improving}, a large-scale text-to-image generation model.

For example, in Fig.~\ref{fig:qual1} (a)-(f), GLIGEN~\cite{li2023gligen} mainly focuses on generating humans and objects, failing to depict the accurate interactions related to the interaction words. 
While InteractDiffusion~\cite{hoe2024interactdiffusion} improves interaction representation, its results show inaccurate or ambiguous interactions that lack both accurate object rendering (e) or fine-grained interaction details (a)-(d) and (f).

In particular, in Fig.\ref{fig:qual1} (g), our model generates a visually enhanced interaction that closely resembles the ground truth interaction region, showing fluid coming out from a bottle. In the multi-humans with one interaction scenario (f), our results depict people holding foods similar to the image from Dall-E 3, while other models only show people sitting at the table. 
This demonstrates that our model further concentrates on detailed interaction regions and distinguishes subtle differences between interaction words with similar spatial arrangements (\eg ``sitting at'' and ``eating at'').
Overall, our model demonstrates a similar level of interaction understandability as a large-scale model such as Dall-E 3 and more accurately exhibits the intended interactions, closely matching with the ground-truth interactions.

We further evaluate our model with complex prompts with multiple interactions to test the ability to distinguish interaction words. 
As shown in Fig.~\ref{fig:teaser}, when given multiple interactions, our model accurately depicts each one, while other models struggle. For example, GLIGEN places objects correctly but fails to represent interactions like ``walking,'' ``jumping,'' and ``throwing.'' InteractDiffusion captures some interactions but misses key objects, like an umbrella. In contrast, our model generates accurate images that reflect the intended interactions. 
This shows that our model successfully captures the semantic difference between interaction words, generating more detailed interactions.
More qualitative results are in the supplementary materials.
% Appendix Sec.~\ref{sec:addition_qualitive}.

\subsection{Quantitative Results}

Tab.~\ref{tab:simscore} compares similarity scores between VerbDiff and previous methods. Our model achieves the highest scores across nearly all evaluation settings.
In the text-to-text (T2T) similarity evaluation, VerbDiff outperforms existing methods, and the difference becomes even more significant in the S-BERT evaluation. 
This highlights that the S-BERT metric better captures the semantic differences within interaction verbs. Although the text-to-image (T2I) score is lower than SD, our model achieves higher scores than the models that leverage additional bounding boxes, demonstrating the ability of our model to understand interactions.
To assess a more comprehensive interaction understanding of VerbDiff, we also report the VQA-Score, which evaluates how well generative models understand more complex compositional prompts, such as interactions. Although our method shows a slight improvement in the I2T and VQA scores compared to SD, it scores significantly higher than InteractDiffusion. 
% This table shows that our model better distinguishes the semantics between verbs and generates high-quality images with intended interactions.

\begin{table}[t]
\centering
\scalebox{0.83}{
\begin{tabular}{l|cc|c|ll}
\hline
\begin{tabular}[c]{@{}c@{}}\\Models\\ \end{tabular} & \multicolumn{2}{c|}{CLIP} & S-BERT         & \multicolumn{2}{c}{VQA-Score}   \\
                                                                     & T2T     & T2I             & T2T            & I2T            & VQA            \\ \hline
HICO-DET & 0.604 & 0.233 & 0.446 & 0.755 & 0.754 \\ \hline
SD~\cite{rombach2022high}                                                                   & {\underline{0.725}}   & \textbf{0.251}  & {\underline{0.620}}          & {\underline{0.769}}          & {\underline{0.765}}          \\
GLIGEN~\cite{li2023gligen}                                                               & 0.683   & 0.238           & 0.554          & 0.679          & 0.674          \\
InteractDiffusion~\cite{hoe2024interactdiffusion}                                                     & 0.703   & 0.232           & 0.575          & 0.728          & 0.734               \\ \hline \hline
VerbDiff (Ours)                                                             & \textbf{0.733}   & {\underline{0.242}}     & \textbf{0.633} & \textbf{0.771} & \textbf{0.766} \\ \hline

\end{tabular}%
}
\caption{\textbf{Similarity comparison between VerbDiff and other models.} We evaluate scores on CLIP, S-BERT~\cite{reimers2019sentence} and a large vision-language alignment benchmark VQA-Score~\cite{lin2025evaluating}.}
\label{tab:simscore}
\end{table}

\renewcommand{\arraystretch}{1.3}
\begin{table}[t]
\centering
\scalebox{0.64}{
\begin{tabular}{l|cccc|cccc}
\hline
\multirow{3}{4em}{Models} & \multicolumn{4}{c|}{SOV-STG-S (Acc $\uparrow$)}               & \multicolumn{4}{c}{SOV-STG-Swin-L (Acc $\uparrow$)}                \\ \cline{2-9} 
                       & \multicolumn{2}{c}{Def.} & \multicolumn{2}{c|}{KO.} & \multicolumn{2}{c}{Def.} & \multicolumn{2}{c}{KO.} \\ \cline{2-9} 
                       & Full        & Rare       & Full        & Rare       & Full        & Rare       & Full       & Rare      \\ \hline
HICO-DET  & 26.52 & 6.78 & 28.68 & 7.29 &29.98 & 12.66 & 31.16 & 13.43 \\ \hline
SD~\cite{rombach2022high}                      & 16.09           & 4.59          & 18.22            & 4.85          & 20.08           & 8.07             & 21.69          & 8.66       \\
GLIGEN~\cite{li2023gligen}                  & 15.88           & 4.85          & 17.91           & 5.24          & 17.83           & 7.00           & 19.35           & 7.57         \\
InteractDiffusion~\cite{hoe2024interactdiffusion}       & \underline{19.67}           & \underline{7.00}          & \underline{21.31}           & \underline{7.69}          & \underline{23.53}           & \underline{10.27}          & \underline{24.86}           & \underline{11.18}          \\ \hline \hline 
VerbDiff (Ours)               & \textbf{22.59}           & \textbf{7.62}          & \textbf{24.79}           & \textbf{7.83}          & \textbf{27.05}            & \textbf{12.60}           & \textbf{28.43}           & \textbf{13.18}          \\ \hline
\end{tabular}%
}
\caption{\textbf{HOI accuracy comparison between VerbDiff and previous methods.} Def. and KO. refer to Default and Known Object.}
\label{tab:HOI_acc}
\end{table}

Tab.~\ref{tab:HOI_acc} presents the HOI accuracy scores, 
where VerbDiff achieves the highest accuracy across all settings compared to previous methods.
In particular, although different backbones are used for computing accuracy, VerbDiff consistently shows the highest accuracy across all settings.
This demonstrates that our method has robust interaction word understanding and produces images with precise interactions. 
Overall, VerbDiff consistently outperforms other models across nearly all evaluation metrics, demonstrating its strong ability to comprehend interaction words and generate high-quality images with accurate human-object interactions, even without extra conditions.
% Overall, VerbDiff consistently outperforms other models across nearly all evaluation metrics, highlighting its strong capability in comprehending interaction words and generating high-quality images with accurate interactions. 
% These results emphasize the comprehensive strength of VerbDiff in effectively capturing and representing complex human-object interactions without extra conditions.

\subsection{Ablation Study}
\begin{figure}
    \centering
    \includegraphics[width=1.0\linewidth]{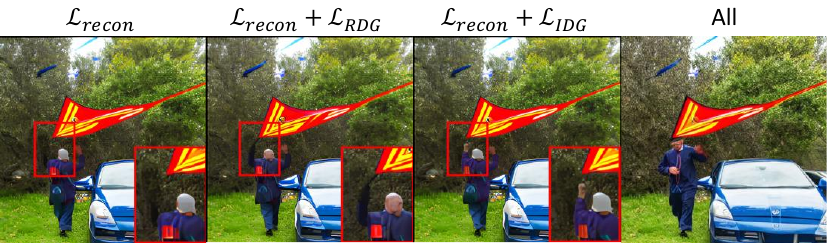}
    \caption{\textbf{Image interaction comparison with and without guidance.} IDG focuses on the specific localized interaction region (red box) in the image.}
    \label{fig:ablationfig1}
\end{figure}

\begin{table}[t]
\centering
\scalebox{0.56}{
\begin{tabular}{c|cccc|c|c|cc}
\hline
\multirow{2}{4em}{VerbDiff} & \multirow{2}{4em}{$\mathcal{L}_{rec}$} & \multirow{2}{4em}{$\mathcal{L}_{triple}$} & \multirow{2}{4em}{$\mathcal{L}_{align}$} & \multirow{2}{4em}{$\mathcal{L}_{IDG}$} & CLIP & S-BERT & \multicolumn{2}{c}{HOI Acc} \\
                        &                           &                            &                           &                         & T2T  & T2T    & Def.          & KO.         \\ \hline
(a)                     & \checkmark                     &                           &                           &                          &     0.691   &  0.582      & 19.38              & 20.89            \\
(b)                     & \checkmark                        & \checkmark                          & \checkmark                         &                         &   0.700      &  0.589      &   20.32            &   21. 87          \\
(c)                     & \checkmark                         &                            &                           & \checkmark                       &  0.699    &    0.588    &       20.21            &    21.67    \\ 
(d)                     & \checkmark                         & \checkmark                            &                      & \checkmark                       &  0.710    &   0.610     &    23.39            &    24.51         \\ \hline \hline
All (Ours)              & \checkmark                         & \checkmark                          & \checkmark                         & \checkmark                       &  \textbf{0.733}   &   \textbf{0.633}     & \textbf{27.05}              & \textbf{28.43}            \\ \hline
\end{tabular}%
}
\caption{\textbf{Ablation of the VerbDiff guidance settings.} We score the similarity and HOI accuracy on the Full setting. We use the SOV-STG-Swin-L model. Combining all the proposed loss functions shows the best performance.}
\label{tab:ablation1}
\end{table}

We show the effectiveness of RDG and IDG in Tab.~\ref{tab:ablation1} and Fig.~\ref{fig:ablationfig1}.
As shown, both RDG (w/ $\mathcal{L}_{triple}$ and $\mathcal{L}_{align}$) (b) and IDG (c) enhance the similarity score and HOI accuracy.
Specifically, (b) scores slightly higher than (c) demonstrating that the model RDG distinguishes the semantics in interaction words. 
While scores improve significantly when using IDG and RDG without $\mathcal{L}_{align}$, the model with all guidance terms achieves the highest overall performance.
In Fig.~\ref{fig:ablationfig1}, we compare images generated with and without each type of guidance. 
Notably, the enlarged interaction region (red box) shows that the model with IDG produces detailed representations (\eg human hands), unlike models with only RDG. This indicates that IDG effectively captures fine-grained details in localized interaction regions.
More ablations are in the supplementary materials. 
% Sec.~\ref{sec:addition_ablation}.

\section{Conclusion}
\label{sec:con}
We propose VerbDiff, a novel text-to-image (T2I) diffusion models that address the interaction misunderstanding problem in SD without additional conditions.
Although previous methods leverage additional conditions to help the model understand the interaction between humans and objects, they still rely on precise instructions and lack an understanding of the semantic differences between interaction verbs.
Our model successfully captures the semantic meanings inherent in interaction words and generates high-quality images with accurate interactions.
Extensive experiments demonstrate the effectiveness of our method in enhancing the ability to understand interaction words of T2I models, achieving better interaction comprehension compared to the previous approaches.

\noindent\textbf{Acknowledgments.} 
This was supported by the Institute of Information \& Communications Technology Planning \& Evaluation (IITP) grant funded by the Korean government(MSIT) (No.RS-2020-II201373, Artificial Intelligence Graduate School Program(Hanyang University)).
{
    \small
    \bibliographystyle{ieeenat_fullname}
    \bibliography{main}
}

\clearpage
\setcounter{page}{1}
\maketitlesupplementary
\appendix

\section{Data Analysis}
\label{sec:data}

\begin{figure}
    \centering
    \includegraphics[width=1.0\linewidth]{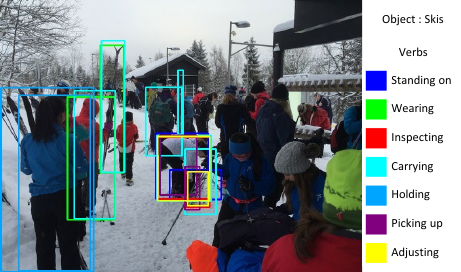}
    \caption{\textbf{An image sample from HICO-DET
    % ~\cite{chao2018learning} 
    containing multiple human-object interactions.} Each colored box corresponds to a distinct human-object interaction, representing different interaction words.}
    \label{fig:multiinteract}
\end{figure}

\begin{figure}
    \centering
    \includegraphics[width=1.0\linewidth]{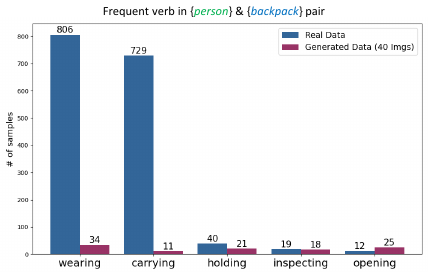}
    \caption{\textbf{Frequent verb in real data and the number of samples contain the frequent verb in captions extracted from generated images.}  
     The blue bars indicate the number of samples in the real dataset for each interaction verb associated with ``backpack''. The purple bars represent the number of captions containing the most frequent verb ``wearing,'' extracted from 40 images generated with ground-truth verbs.}
    \label{fig:freqeuntverb}
\end{figure}

\subsection{HICO-DET Dataset}
We present an example from HICO-DET
% ~\cite{chao2018learning} 
containing multiple interactions within a single image. As shown in Fig.~\ref{fig:multiinteract}, the image includes various interaction verbs (\eg, ``wearing''), which can confuse the model when trained using the conventional reconstruction loss. To isolate interactions corresponding to the correct interaction verbs, we apply a mask $\mathcal{M}$ during training.

\subsection{Anchor Interaction Words}
Fig.~\ref{fig:freqeuntverb} presents two graphs for the human and ``backpack'' pair example. We generate 40 images per prompt (\eg, ``A person \{verb\} a backpack'') and extract captions using InstructBLIP.
% \cite{liu2024visual}. 
We then count the captions that include the most frequent verb from the real data (\eg, ``wearing''). Across all generated samples, nearly half of the captions contain the frequent verb, which we define as the anchor interaction word for relation disentanglement guidance.

\section{Implementation $\And$ Evaluation Details}
\subsection{Implementation Details}
\label{sec:implement}
We use ``black and white image, extra arms, extra legs'' as negative prompts when generating biased interaction images during training. Additionally, we include ``naked, poor resolution'' to negative prompts for image generation in all our experiments. For encoding texts and images, we utilize CLIP ViT-L/14 weights, consistent with the weights used for the text encoder in SD.
% ~\cite{rombach2022high}.

\subsection{Sentence-BERT Evaluation Metric}
\label{sec:sbert}

\begin{table}[t]
\centering
\scalebox{0.83}{%
\begin{tabular}{|cc|cc|}
\hline
\multicolumn{2}{|c|}{CLIP}                                             & \multicolumn{2}{c|}{S-BERT}                       \\ \hline
\multicolumn{1}{|c|}{Verb}                & \multicolumn{1}{c|}{Score} & \multicolumn{1}{c|}{Verb}                & Score  \\ \hline
\multicolumn{1}{|c|}{\textbf{riding}}     & 1.0 
& \multicolumn{1}{c|}{\textbf{riding}}     & 1.0    \\ \hline
\multicolumn{1}{|c|}{\textbf{hopping}}    & 0.9595                     & \multicolumn{1}{c|}{\textbf{sitting on}} & 0.8944 \\ \hline
\multicolumn{1}{|c|}{\textbf{sitting on}} & 0.9585                     & \multicolumn{1}{c|}{\textbf{straddling}} & 0.8385 \\ \hline
\multicolumn{1}{|c|}{\textbf{straddling}} & 0.9551                     & \multicolumn{1}{c|}{holding}             & 0.8745 \\ \hline
\multicolumn{1}{|c|}{\textbf{walking}}    & 0.9512
& \multicolumn{1}{c|}{\textbf{walking}}    & 0.8711 \\ \hline
\multicolumn{1}{|c|}{carrying}            & 0.9341                     & \multicolumn{1}{c|}{carrying}            & 0.8555 \\ \hline
\multicolumn{1}{|c|}{pushing}             & 0.9277                     & \multicolumn{1}{c|}{jumping}             & 0.8468 \\ \hline
\multicolumn{1}{|c|}{jumping}             & 0.9111                     & \multicolumn{1}{c|}{\textbf{hopping}}    & 0.8436 \\ \hline
\multicolumn{1}{|c|}{holding}             & 0.8960                     & \multicolumn{1}{c|}{parking}             & 0.8218 \\ \hline
\multicolumn{1}{|c|}{parking}             & 0.8950                     & \multicolumn{1}{c|}{inspecting}          & 0.7982 \\ \hline
\multicolumn{1}{|c|}{inspecting}          & 0.8662                     & \multicolumn{1}{c|}{pushing}             & 0.7975 \\ \hline
\multicolumn{1}{|c|}{repairing}           & 0.8291                     & \multicolumn{1}{c|}{repairing}           & 0.7456 \\ \hline
\multicolumn{1}{|c|}{washing}             & 0.8081                     & \multicolumn{1}{c|}{washing}             & 0.7253 \\ \hline
\end{tabular}%
}
\caption{\textbf{Cosine similarity comparison between CLIP
% ~\cite{radford2021learning}
and Sentence-BERT (S-BERT).
% ~\cite{reimers2019sentence}.
} We score the similarity of the verb ``riding'' with other verbs associated with the object ``bicycle'' in real data.}
\label{tab:sbert_supple}
\end{table}

We compare the cosine similarity differences between CLIP and Sentence-BERT, focusing on interaction verbs associated with the object ``bicycle''. Using the text template: ``A photo of a person \{verb\} a bicycle,'' we calculate cosine similarities, as shown in Tab.~\ref{tab:sbert_supple}. As highlighted by the scores for bolded verbs (\eg, ``walking''), CLIP evaluates all verbs as highly similar, even when humans perceive them as distinct. In contrast, S-BERT successfully distinguishes between interaction verbs, better reflecting the interaction differences that humans recognize in real human-object interactions.

\begin{figure}[t]
    \centering
    \includegraphics[width=1.0\linewidth]{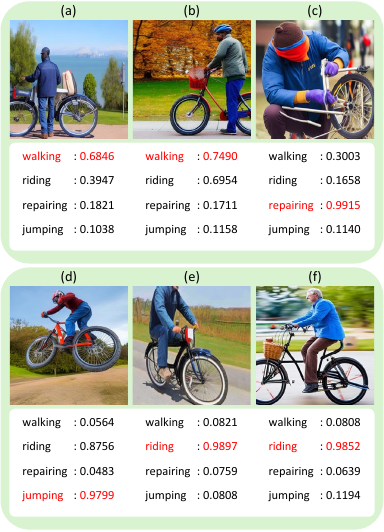}
    \caption{\textbf{VQA-score
    % ~\cite{lin2025evaluating} 
    result examples.}  
    We measure the probability that the VQA model answers ``yes'' for questions based on four verbs associated with the ``bicycle'' class. The verb with the highest score is highlighted in red.}
    \label{fig:vqa_supp}
\end{figure}

\subsection{VQA-Score}
\label{sec:VQA}
For assessing the image-to-text (I2T) alignment score and VQA-score, we use CLIP-FlanT5 weights. Additionally, we apply the following question template: ``Is this figure showing a \{H\} \{R\} a/an \{O\}? Please answer yes or no''. Fig.~\ref{fig:vqa_supp} illustrates VQA-score examples for the ``bicycle'' class. We use four verbs in the question template and calculate the probability scores for each image. As shown, the VQA-score effectively distinguishes complex prompts involving human-object interactions.
 In particular, the model differentiates images with ambiguous interactions (b), successfully identifying the human foot on the ground.

\section{Additional Ablations}
\label{sec:addition_ablation}
\subsection{Model Weight Comparison}

\begin{table}[t]
\centering
\scalebox{0.83}{%
\begin{tabular}{l|c|c|l|c|c}
\hline
\multicolumn{1}{c|}{$\mathcal{L}_{IDG}$} & CLIP                       & S-BERT                     & \multicolumn{1}{c|}{$\mathcal{L}_{rec}$} & CLIP                       & S-BERT                    \\ \hline
SD                                      & \multicolumn{1}{c|}{0.725} & \multicolumn{1}{c|}{0.620} & SD                                      & \multicolumn{1}{c|}{0.725} & \multicolumn{1}{c}{0.620} \\ \hline
0.2                                      & 0.725                      & 0.638                      & 0.4                                      & 0.733                      & 0.634                     \\
0.4                                      & 0.723                      & 0.638                      & 0.7                                      & 0.732                      & 0.636                     \\
0.8                                      & \textbf{0.729}             & \textbf{0.640}             & 1.0                                      & \textbf{0.735}             & \textbf{0.638}            \\
1.2                                      & 0.726                      & 0.639                      & 1.5                                      & 0.732                      & 0.637                     \\
1.6                                      & 0.724                      & 0.639                      & 2.0                                      & 0.731                           &0.636                           \\ \hline
\end{tabular}%
}
\caption{\textbf{Score variations across the interaction direction guidance (IDG) and reconstruction loss weights.} Scores are evaluated using CLIP and Sentence-BERT (S-BERT) text-to-text similarity metrics. The top row represents the scores from Stable Diffusion (SD).}
% ~\cite{rombach2022high}.}
\label{tab:IDG_recon_supp}
\end{table}

\begin{table}[t]
\centering
\scalebox{0.83}{%
\begin{tabular}{l|c|c|l|c|c}
\hline
\multicolumn{1}{c|}{$\mathcal{L}_{RDG}$} & CLIP           & S-BERT         & \multicolumn{1}{c|}{$m$} & CLIP           & S-BERT         \\ \hline
SD                                     & 0.725          & 0.620          & SD& 0.725          & 0.620          \\ \hline
1                                        & 0.735          & 0.640          & 0.1                         & 0.733          & 0.634          \\
5                                        & 0.735          & 0.641          & 0.2                         & \textbf{0.735}          & \textbf{0.638}\\
10                                       & \textbf{0.736} & \textbf{0.641} & 0.4                         & 0.732 & 0.636 \\
20                                       & 0.734          & 0.640          & \multicolumn{1}{c|}{-}      & -              & -              \\ \hline
\end{tabular}%
}
\caption{\textbf{Score variations based on the weight of relation disentanglement guidance (RDG) and the margin $m$ in $\mathcal{L}_{triple}$ for RDG.} Scores are evaluated using CLIP and Sentence-BERT (S-BERT) text-to-text similarity metrics. The top row represents the scores from Stable Diffusion (SD).}
% ~\cite{rombach2022high}.}
\label{tab:RDG_supp}
\end{table}

\begin{table}[t]
\centering
\scalebox{0.83}{%
\begin{tabular}{l|cccc}
\hline
\multirow{3}{*}{Model}    & \multicolumn{4}{c}{SOV-STG-S (Acc $\uparrow$)}                                        \\ \cline{2-5} 
                          & \multicolumn{2}{c}{Def.}                  & \multicolumn{2}{c}{KO.}        \\ \cline{2-5} 
                          & Full           & \multicolumn{1}{c}{Rare} & Full           & Rare          \\ \hline
SD                        & 16.09          & 4.59                      & 18.22          & 4.85          \\ \hline
w/o $\alpha$ & 17.83              & 5.48                         & 19.12              & 5.65             \\
w/ \;\,$\alpha$   & \textbf{22.59} & \textbf{7.62}             & \textbf{24.79} & \textbf{7.83} \\ \hline
\end{tabular}%
}
\caption{\textbf{HOI accuracy comparison with and without adaptive interaction modification $\alpha$.} Scores are evaluated using SOV-STG-S
% ~\cite{chen2023focusing}
weights. The top row represents the results from Stable Diffusion (SD).}
% ~\cite{rombach2022high}.}
\label{tab:CB_comparison}
\end{table}

We compare the weight hyperparameters for relation disentanglement guidance (RDG), interaction direction guidance (IDG), reconstruction loss, and the margin value $m$ in  $\mathcal{L}_{triple}$ of RDG in Tab.~\ref{tab:RDG_supp} and Tab.~\ref{tab:IDG_recon_supp}, respectively. Additionally, we include the second-best model (SD) in the top row of each table for reference. The overall scores differ from the main evaluation because we evaluate in different settings to observe score differences more clearly.

Tab.~\ref{tab:CB_comparison} shows the HOI accuracy differences when using adaptive interaction modification  $\alpha$. To highlight the effectiveness of $\alpha$ in balancing modification across interaction words, we compare it under the same settings as the main evaluation table. The results demonstrate that adaptive interaction modification successfully balances the extent of interaction adjustments, preventing overfitting interaction words with many samples in the real dataset.

\subsection{Self-Attention $\And$ Cross-Attention}

\begin{figure}
    \centering
    \includegraphics[width=1.0\linewidth]{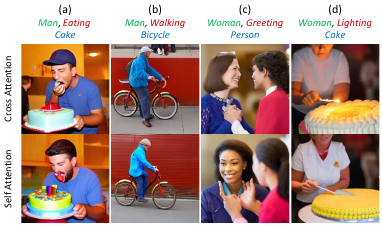}
    \caption{\textbf{Comparison between self-attention and cross-attention layer tuning.} While the cross-attention-tuned model generates images with accurate interactions, the self-attention-tuned model fails to capture precise interactions.}
    \label{fig:selfcross_sup}
\end{figure}

We show the interaction differences depending on which layer is tuned in Fig.~\ref{fig:selfcross_sup}. As shown, the self-attention-tuned model fails to understand the semantics of interaction verbs and generates images with inaccurate interactions. However, the cross-attention tuned model (VerbDiff) accurately depicts the intended interactions. This demonstrates that the cross-attention layer better reflects nuanced interactions, enhancing the interaction word understanding in SD.

\subsection{Interaction Region}
\begin{figure}[!t]
    \centering
    % \vspace{-2mm}
    \includegraphics[width=1.0\linewidth]{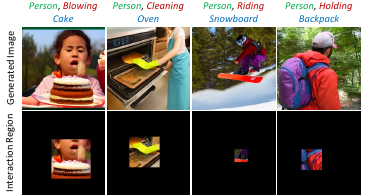}
    \caption{\textbf{Interaction region in generated images.}}
    \vspace{-2mm}
    \label{fig:interactregion}
\end{figure}

We present the generated images and the extracted interaction regions utilized during training in Fig.~\ref{fig:interactregion}.
As can be seen, the IR module extract quite accurate interaction region between humans and objects.

% \vspace{-2mm}

\section{Additional Qualitative Results}
\label{sec:addition_qualitive}

\begin{figure*}[t]
    \centering
    \includegraphics[width=0.9\textwidth, height=0.8\textwidth]{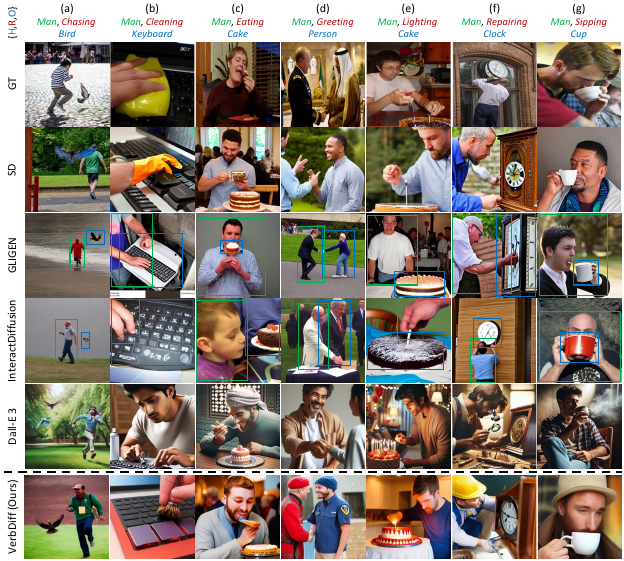}
    \caption{\textbf{Additional single interaction qualitative results.} The colored boxes mean the input bounding boxes.}
    \label{fig:qualsupp}
\end{figure*}

Fig.~\ref{fig:qualsupp} presents additional results comparing VerbDiff with SD,
% ~\cite{rombach2022high},
GLIGEN,
% ~\cite{li2023gligen}, 
InteractDiffusion,
% ~\cite{hoe2024interactdiffusion}, 
% and the large-scale text-to-image generation model DALL-E. 3~\cite{betker2023improving}.
Our model generates high-quality images with accurate interactions that closely resemble the ground truth and those produced by DALL-E 3. For example, in Fig.~\ref{fig:qualsupp} (e), it successfully captures a man lighting a candle, while GLIGEN fails to generate the candle, and InteractDiffusion does not depict the action correctly.

\begin{figure*}[t]
    \centering
    \includegraphics[width=0.85\textwidth, height=0.95\textwidth]{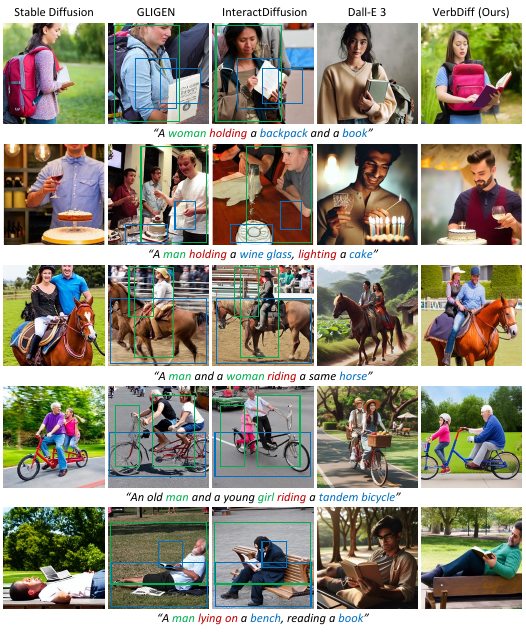}
    \caption{\textbf{Additional multiple interactions qualitative results.} The colored boxes represent the corresponding human and object bounding boxes. We extract the box from our generated images and apply it to the GLIGEN
    % ~\cite{li2023gligen} 
    and InteractDiffusion.}
    % ~\cite{hoe2024interactdiffusion}
    \label{fig:multisupp}
\end{figure*}

Additionally, Fig.~\ref{fig:multisupp} shows diverse images from complex prompts with multiple interactions. Even when provided with precise bounding boxes extracted from our generated images, GLIGEN and InteractDiffusion fail to produce the intended interactions or objects accurately. In contrast, our model captures interactions effectively without additional conditions, achieving quality comparable to DALL-E 3 and enhancing interaction understanding within SD.
% Our model generates higher-quality images with accurate interactions that closely resemble the ground-truth interactions and those produced by DALL-E 3. For example, in Fig.~\ref{fig:qualsupp} (e), our model successfully captures the precise interaction of a man lighting a candle, closely mirroring the ground truth and DALL-E 3 images. In contrast, GLIGEN fails to generate a candle on the cake, and InteractDiffusion does not depict the lighting action correctly.

% Additionally, Fig.~\ref{fig:multisupp} shows diverse generated images using complex prompts with multiple interactions. We extract bounding boxes from our generated images and provide them as additional grounding information for GLIGEN and InteractDiffusion. Despite this, both models fail to produce the intended interactions or objects accurately, even with precise bounding boxes.
% In comparison to SD, GLIGEN, and InteractDiffusion, our model effectively captures the intended interactions without requiring additional conditions. It achieves high-quality images comparable to those generated by DALL-E 3. This demonstrates that our model accurately matches intended interactions with precise interaction words, significantly enhancing interaction understanding within SD.

\section{Limitations}

\begin{figure}[!t]
    \centering
    \includegraphics[width=1.0\linewidth]{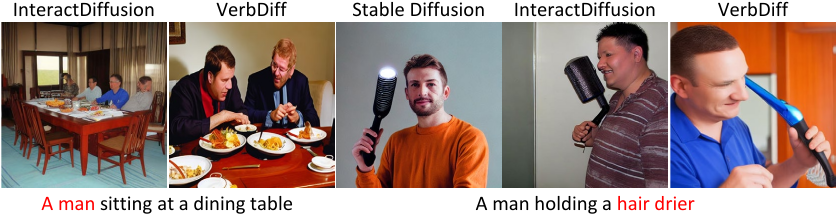}
    \vspace{-4mm}
    \caption{\textbf{Failure Cases with previous methods.}}
    \label{fig:failure}
\end{figure}

VerbDiff generates high-quality images with accurate interactions by improving the interaction word comprehension in SD. 
While our model achieves better scores across various evaluation metrics, there are cases where several humans perform the same interaction within a single training image. 
As shown in Fig.~\ref{fig:failure}, VerbDiff occasionally miscounts the number of humans and inherits the limitations of the Stable Diffusion (SD) framework.

Additionally, it still produces some deviations from real interactions when handling complex prompts involving multiple human-object interactions. 
This limitation may arise from the model focusing on distinguishing interaction words without considering similarities between interactions involving different objects (\eg, ``walking a bicycle'' and ``walking a motorcycle’’). We anticipate that incorporating more generalized representations of interactions between humans and objects could further enhance interaction comprehension in text-to-image models.

%여기에 multiple 이랑 좀 더 generaal 한 interaction 고려할 필요가 있다는 식으로 쓰면서 마무리 

\end{document}